\documentclass{article}
            \usepackage{graphicx}
            \usepackage{amsmath,amssymb}
            \begin{document}
            \title{New Gauge Field from Extension of
            Space Time Parallel Transport of
            Vector Spaces to the Underlying Number Systems}
            \author{Paul Benioff\\
            Physics Division, Argonne National
            Laboratory,\\ Argonne, IL 60439, USA\\
            e-mail: pbenioff@anl.gov}

            \maketitle

            \begin{abstract}
            One way of describing gauge theories in physics is to
            assign a vector space $\bar{V}_{x}$ to each space
            time point $x.$ For each $x$ the field $\psi$ takes
            values $\psi(x)$ in $\bar{V}_{x}.$ The freedom to
            choose a basis in each $\bar{V}_{x}$ introduces gauge group
            operators and their Lie algebra representations to
            define parallel transformations between vector spaces.
            This paper is an exploration of the extension of
            these ideas to include the underlying
            scalar complex number fields. Here a Hilbert space,
            $\bar{H}_{x},$ as an example of $\bar{V}_{x},$ and a
            complex number field, $\bar{C}_{x},$ are associated with
            each space time point. The freedom to choose a basis in
            $\bar{H}_{x}$ is expanded to include the freedom to choose
            complex number fields. This expansion is based on the
            discovery that there exist representations of complex
            (and other) number systems that differ by arbitrary
            scale factors. Compensating changes must be made in the
            basic field operations so that the relevant axioms are
            satisfied. This results in the presence of a new real valued gauge
            field $\vec{A}(x).$ Inclusion of $\vec{A}(x)$ into
            covariant derivatives in Lagrangians  results in the
            description of $\vec{A}(x)$ as a gauge boson that can have
            mass. The great accuracy of QED suggests that the
            coupling constant of $\vec{A}(x)$ to matter fields is
            very small compared to the fine structure constant.
            Other physical properties of $\vec{A}(x)$ are
            not known at present.
            \end{abstract}

            \section{Introduction}
            One approach to the description of some physical theories
            is based on the assignment of a vector space to each
            space time point. Gauge theories are examples of these
            theories. The usefulness of this approach and the
            resultant freedom of basis choice in the vector spaces
            \cite{Yang} has resulted in the creation of several
            different gauge theories. Included are Quantum
            electrodynamics, quantum chromodynamics, and the
            standard model \cite{Glashow,Weinberg,Salam,Novaes}.

            In this approach to gauge theories \cite{Mack,Montvay}
            there is just one set, $\bar{C},$ of
            complex numbers that is the common scalar field for all
            the vector spaces, $\bar{V}_{x}.$  The scalars in
            scalar-vector multiplication and scalar products of
            vectors in $\bar{V}_{x}$ take values in $\bar{C}$
            independent of $x.$

            The purpose of this paper is to explore some consequences of
            expanding the usual setup by replacing $\bar{C}$ with different
            complex number structures $\bar{C}_{x}.$ In this case a pair
            $\bar{V}_{x},\bar{C}_{x}$ is associated with each space time
            point $x.$ The freedom of choice of basis sets in each $\bar{V}_{x}$
            \cite{Yang} is expanded here to include freedom of choice of
            the complex number structures $\bar{C}_{x}.$

            The replacement of one common complex number structure,
            $\bar{C},$ with structures $\bar{C}_{x}$ at each point
            $x,$ affects nonlocal functions such as space time
            derivatives and integrals. An example is the
            derivative, $\partial_{\mu,x},$ in direction
            $\mu,$ of a complex valued field $\psi(x)$ where $\psi(x)$
            is a number value in $\bar{C}_{x}.$  The derivative is given
            by $\partial_{\mu,x}$ as\begin{equation}\label{dmux}
            \partial_{\mu,x}\psi=\frac{\psi(x+dx^{\mu})-\psi(x)}{dx^{\mu}}.
            \end{equation}

            There are two problems with this expression.  One is a
            consequence of the fact that $\psi(x+dx^{\mu})$ and
            $\psi(x)$ are in different complex number structures,
            and subtraction is not defined between
            elements of different structures. It is defined only
            within a structure. The other is that the
            "naheinformationsprinzip" \cite{Mack,Montvay} "no
            information at a distance"  principle forbids access, at
            $x,$ to a number value in a structure at a different site.
            Thus $\psi(x+dx^{\mu})$ is not available to an observer at $x$
            with structure $\bar{C}_{x}.$

            One way to solve these problems is to replace
            $\psi(x+dx^{\mu})$ by $\psi(x+dx^{\mu})_{x}$  where
            $\psi(x+dx^{\mu})_{x}$ is the same number in
            $\bar{C}_{x}$ as $\psi(x+dx^{\mu})$ is in $\bar{C}_{x+dx^{\mu}}.$
            $\bar{C}_{x+dx^{\mu}}$ is the complex number
            structure at point $x+dx^{\mu}.$
            In this case $\partial_{\mu,x}\psi$ becomes\begin{equation}
            \label{dmuxp}\partial_{\mu,x}\psi=\frac{\psi(x+
            dx^{\mu})_{x}-\psi(x)}{dx^{\mu}}.\end{equation} Use of
            this expression in physical theories involving space
            time derivatives gives the same results as does use of
            Eq. \ref{dmux}. This would suggest that nothing is to be
            gained from replacing one $\bar{C}$ everywhere with
            $\bar{C}_{x}$ at each point $x.$

            The realization that, for each type of number,
            there exists an infinite number of different representations
            that differ from one another by scaling factors, makes
            possible a generalization of the above.
            One way to proceed is to define for each $\mu$
            a complex number structure $\bar{C}^{r}_{x}$ that is the
            local representation of $\bar{C}_{x+dx^{\mu}}$ on $\bar{C}_{x}.$

            $\bar{C}^{r}_{x}$ is related to
            $\bar{C}_{x}$ through a scaling factor $r=r_{\mu,x}.$
            Here $r_{\mu,x}$ is a real number in $\bar{C}_{x}$ that
            is the $\mu$ component of $r_{y,x}$ which is associated with
            the link from $x$ to $y$ where $y=x+\hat{\nu}dx.$ The
            relation between $\bar{C}^{r}_{x}$ and $\bar{C}_{x}$ is
            shown by noting that the number $a$ in $\bar{C}^{r}_{x}$
            \emph{corresponds} to the number $r_{\mu,x}a$ in
            $\bar{C}_{x}.$

            The meaning of correspondence is based on the fact that
            both $\bar{C}^{r}_{x}$ and $\bar{C}_{x}$ are complex
            number structures over the same base set $C.$ The
            overlines,  on $\bar{C}_{x},$ and $\bar{C}^{r}_{x}$ denote
            that they are complex number structures. $C$ without
            an overline denotes a base set. One says that the number value
            $a$ in $\bar{C}^{r}_{x}$ corresponds to the number value $b$
            in $\bar{C}_{x}$ if the element of $C$ that has
            value $a$ in $\bar{C}^{r}_{x}$ has value $b$ in
            $\bar{C}_{x}.$ In the case at hand $b=r_{\mu,x}a.$
            Note that the element of $C$ that has
            value $a$ in $\bar{C}^{r}_{x}$ is different from the
            element of $C$ that has the same value $a$ in
            $\bar{C}_{x}.$

            The mathematical logical \cite{Barwise,Keisler}
            description of mathematical systems, as a structures consisting
            of a base set, basic operations and relations, and
            constants is used in the above. The structures are required to
            satisfy an appropriate set of axioms. Both $\bar{C}^{r}_{x}$
            and $\bar{C}_{x}$ are structures, on $C$, that satisfy
            the complex number axioms. As a result, one structure is just
            as valid as the other and either one can serve as a complex
            number base in physics. This is the case even though
           the representations of the  basic operations in $\bar{C}^{r}_{x},$
           include the scale factor and the basic operations of
           $\bar{C}_{x}.$ This equal validity of the structures, as
           complex number systems, is fundamental to this paper.

            For the following a gauge field representation of $r_{y,x}$
            as\begin{equation}\label{rvecA}r_{y,x}=e^{\vec{A}(x)
            \cdot\hat{\nu}dx}=e^{\sum_{\mu}A_{\mu}(x)dx^{\mu}}
            \end{equation} is used. $\vec{A}(x)$ is a
            real valued gauge field with components $A_{\mu}(x)$
            where $r_{\mu,x}=e^{A_{\mu}(x)dx^{\mu}}.$

            The replacement of $\psi(x+dx^{\mu})$ by $r_{\mu,x}\psi(x+
            dx^{\mu})_{x}$ in Eq. \ref{dmux} gives\begin{equation}
            \label{Dmuxp}D_{\mu,x}\psi=\frac{r_{\mu,x}\psi(x+dx^{\mu})_{x}
            -\psi(x)}{dx^{\mu}}.\end{equation} Here $r_{\mu,x}\psi(x
            +dx^{\mu})_{x}$ is a number in $\bar{C}_{x}$ that corresponds
            to a number $\psi(x+dx^{\mu})^{r}_{x}$ in $\bar{C}^{r}_{x}.$
            $\psi(x+dx^{\mu})^{r}_{x}$ is the same number value in
            $\bar{C}^{r}_{x}$ as $\psi(x+dx^{\mu})$ is in
            $\bar{C}_{x+dx^{\mu}}.$

            The inclusion of the factor $r_{\mu,x},$ or its gauge
            field equivalent, into derivatives,  as in $D_{\mu,x}\psi$
            represents one way of including the freedom of choice
            of complex number structures into gauge theories. In
            this case the gauge groups include a factor $GL(1,R)$
            for the gauge field $\vec{A}(x).$ Including this into
            the covariant derivatives in Lagrangians gives the result that
            $\vec{A}(x)$ is a gauge boson for which the presence of a
            mass term in the Lagrangian is optional. Also the great
            accuracy of QED implies that the coupling constant of
            $\vec{A}(x)$ to matter fields must be very small compared
            to the fine structure constant. Other physical properties,
            if any, must await further work.

            It should be noted that the setup described here is a
            generalization of the usual case. To see this, set
            $\vec{A}(x)=0$  for all $x.$ Then the notions of
            "correspondence" and "same number as" coincide.
            $\bar{C}^{r}_{x}$ becomes identical to $\bar{C}_{x},$
            and $D_{\mu,x}$ in Eq. \ref{Dmuxp} becomes $\partial_{
            \mu,x}$ in Eq. \ref{dmuxp}. In this case the different
            $\bar{C}_{x}$ become identical to one another and the
            usual case of one complex number structure, $\bar{C},$
            for all space time points is recovered.

            At present it is not known if physics makes use of this
            generalization. The fact that physics does make use of
            the freedom of basis choice in vector spaces makes it
            reasonable to entertain the possibility that physics
            might make use of the freedom of choice of complex
            number structures as scalars for the vector spaces.

            In any case the purpose of this paper is to explore
            some consequences of the freedom of choice of complex
            number structures as scalars. Physical justification
            of this approach is work for the future.

            This brief summary is expanded, with additional details given in the
            rest of the paper. The space time field of complex numbers with a
            complex number structure, $\bar{C}_{x},$ at each point $x$ is
            described in the next section. Relations between complex
            number structures and their elements at point $y,$ and their
            local representations at point $x$ are discussed.

            Section \ref{GF} describes the gauge field representation
            of $r_{y,x}$ as in Eq. \ref{rvecA}. This is is followed
            by discussions  of path integrals of the gauge field
            and of  space time derivatives and integrals.

            The flexibility of number structures affects other
            mathematical systems that are based on numbers.
            An example is discussed in Section \ref{HSSF} where
            emphasis is placed on Hilbert spaces as examples of
            vector spaces.  Each point $x$ has an associated pair
            $\bar{H}_{x},\bar{C}_{x}.$ The changes in the scalars
            arising from multiplication by $r_{y,x}$ induce
            corresponding changes in the basic operations
            involving scalars that are part of the Hilbert
            space structure. The changes must be such that
            the validity of the Hilbert space axioms
            \cite{Kadison1} is preserved under the change.

            Both Abelian, $U(1),$ and nonabelian, $SU(2),$ gauge
            theories are discussed in Section  \ref{GT}. The main
            difference from the usual description is the expansion
            of the gauge group from $U(n)$ to $GL(1,R)\times U(n).$
             As noted, $\vec{A}(x)$ appears as a gauge boson for
             which a mass term in the Lagrangian is optional.

            The final section \ref{D} is  a discussion,
            mainly of some open questions generated by this work.
            The main ones concern the physical nature, if any,
            of  $\vec{A}(x),$ and and its integrability.

             \section{Field  of Complex Number Structures}
            \label{FCNS}
            The representation of mathematical systems as mathematical structures
            is a basic tenet of mathematical logic \cite{Barwise,Keisler}.
            The usefulness of mathematical structures has also been
            noted by \cite{Tegmark}-\cite{Welch}. (See also
            \cite{Chihara,Shapiro}.) This applies
            to all types of numbers, such as the natural numbers, the
            integers, the rational numbers, the real numbers, and the
            complex numbers.

            The view of each type of number as  structures
            emphasizes the basic  operations and relations along with the base
            set appropriate for each type. The basic relations and operations
            are required to satisfy a set of axioms appropriate for the type
            being considered. For example, the real numbers satisfy axioms for
            a complete ordered field\footnote{A field is a system that is
            closed under addition, subtraction, multiplication, and division. Additive
            and multiplicative identities exist. The relations are associative,
            commutative, and multiplication is distributive over addition.}
            \cite{Rudin}. Complex numbers satisfy the axioms for an
            algebraically closed field
            of characteristic $0$.\footnote{An algebraically closed field is a
            field in which all polynomial equations have solutions in the field.
            Characteristic $0$ means that  $1+1+\cdots,+1\neq 0$ holds for all
            finite strings of ones.} \cite{Adamson}. Because of its usefulness,
            the complex conjugation operation has been added as a basic operation.
            The associated axioms are given in \cite{comcomj}.

            These ideas are used  here to describe a field\footnote{Note the
            different meanings of field appearing here.} of complex number structures
            where a complex number structure, $\bar{C}_{x},$ is associated with each
            point $x$ in $3+1$ dimensional space time, $R^{4}$. The main task is to
            determine the relationship between complex number structures at
            different points in $R^{4}.$

            Let $\bar{C}_{x}$ and $\bar{C}_{y}$ be complex
            number fields associated with  points $x$ and $y.$
            Here $\bar{C}_{x}$ and $\bar{C}_{y}$ are mathematical
            structures denoted by\begin{equation}\label{CxCy}
            \begin{array}{c}\bar{C}_{x}=\{C,\pm_{x},\times_{x},
            \div_{x},^{*_{x}},0_{x},1_{x}\}\\\\\bar{C}_{y}=\{C,\pm_{y},
            \times_{y}, \div_{y},^{*_{y}},0_{y},1_{y}\}.\end{array}
            \end{equation} $C$ is the underlying sets on which the
            structures are defined. As noted in the
            introduction, $C$ without an over line  denotes
            a set. $C$ with an over line, as in $\bar{C},$
            denotes a complex number structure on $C.$ Numbers in
            $\bar{C}_{y}$ and $\bar{C}_{x}$ are denoted with subscripts,
            as in $a_{y},a_{x}.$

            Use of the same underlying set, $C,$ in both $\bar{C}_{y}$ and
            $\bar{C}_{x},$ instead of distinct sets, $C_{y}$ and $C_{x},$
            is not necessary. However, it simplifies the description and
            causes no problems.

            One would like to be able to directly compare the values of
            numbers in $\bar{C}_{y}$ with the values of numbers in $\bar{C}_{x}.$
            Such comparisons occur in space time derivatives where a number
            value in $\bar{C}_{x}$ is subtracted from a number value in
            $\bar{C}_{y}$ with $y$ a neighbor point of $x,$ as in Eq.\ref{dmux}.

            However this is not possible for two reasons.  One is that subtraction
            of number values in different structures is not defined. Subtraction
            and other operations are defined only within structures. They are
            not defined between different structures. The other reason is that the
            "naheinformationsprinzip"\cite{Mack,Montvay} "no
            information at a distance" principle forbids direct
            access to the numbers and their values in $\bar{C}_{y}$  by an
            observer at site $x.$

            The solution to this problem requires that one have available,
            at $x,$ a complex number structure that is
            a local representation of $\bar{C}_{y}$ on $\bar{C}_{x}.$ This is a
            representation of the basic operations, relations, and constants of
            $\bar{C}_{y}$ in terms of the operations, relations, and constants in
            $\bar{C}_{x}.$ This enables a direct determination of the correspondence
            between number values in the two structures. If $a_{y}$ is a number
            value in $\bar{C}_{y},$ the representation gives the number value
            in $\bar{C}_{x}$ that corresponds to $a_{y}$.

            One solution is to simply require that the local representation
            of $\bar{C}_{y}$ on $\bar{C}_{x}$ is $\bar{C}_{x}$ itself. In this
            case, the local representation of a number value, $a_{y}$ in
            $\bar{C}_{x}$ is the number value $a_{x}$, which is the same
            value in $\bar{C}_{x}$ as $a_{y}$ is in $\bar{C}_{y}.$

            However, it turns out that this is unnecessarily restrictive as it
            excludes an infinite number of other possibilities. These are based
            on the discovery that it is possible to define an infinite number of
            different structures of complex numbers, or of any other type of number,
            that differ from one another by scaling factors. The scaling of the
            numbers in the different structures must be compensated for by
            changes in the basic operations and relations  in such a manner that,
            for any pair of structures, one satisfies the complex
            number axioms if and only if the other one does.

             \subsection{The Representation of $\bar{C}_{y}$ on $\bar{C}_{x}$}
            \label{RCyCx}These possibilities are taken account of here
            by letting the local
            representation of $\bar{C}_{y}$ on $\bar{C}_{x}$ differ from
            $\bar{C}_{x}$ by a scaling factor that depends on the link between
            $x$ and $y.$ To see how this works,  let $y$ be a neighbor point of $x.$
            The representation, $\bar{C}^{r_{y,x}}_{x},$ of $\bar{C}_{y}$
            on $\bar{C}_{x}$ is defined to be a complex number structure
            on the  same base set $C,$ (no over line) as is used for
            $\bar{C}_{x}.$ As a structure, $\bar{C}^{r_{y,x}}_{x}$ is
            given by \begin{equation}\label{Ccxatx}\bar{C}^{r_{y,x}}_{x}
            =\{C_{x},\pm^{r_{y,x}}_{x},\times^{r_{y,x}}_{x},\div^{r_{y,x}}_{x},
            \mbox{}^{*^{r_{y,x}}_{x}},0^{r_{y,x}}_{x},1^{r_{y,x}}_{x}\}.
            \end{equation}

            In this definition $r_{y,x}$ is a positive real number value
            associated with the link from $x$ to $y.$
            The righthand subscript in $r_{y,x}$ denotes the complex number
            field to which it belongs. Thus $r_{y,x}$
            is an element of $\bar{C}_{x}$ and $r_{x,y}$ is an element of
            $\bar{C}_{y}.$ The order, $y,x,$  of the subscripts
            shows that $r_{y,x}$ is associated with the link from
            $x$ to $y$ and $r_{x,y}$ is associated with the same
            link in the opposite direction. Also, to save on notation, $r$
            is often used as a short representation of $r_{y,x}.$

            The three structures, $\bar{C}_{y},$ $\bar{C}_{x}^{r}$,
            and $\bar{C}_{x}$ can be isomorphically mapped into one
            another by the use of two isomorphisms,   $W^{y}_{r}$  and
            $W^{r}_{x},$ where\begin{equation}\label{FxyWx} \bar{C}_{y}
            =W^{y}_{r}\bar{C}^{r}_{x}=W^{y}_{r}W^{r}_{x}\bar{C}_{x}
            =F_{y,x}\bar{C}_{x}.\end{equation}
            $W^{r}_{x}$ and $W^{y}_{r}$ are isomorphisms in
            that $W^{y}_{r}$ satisfies \begin{equation}
            \label{Rppropyxx}\begin{array}{c}W^{y}_{r}(a^{r}_{x})=a_{y},
            \\\\W^{y}_{r}(a^{r}_{x}O^{r}_{x}b^{r}_{x})=
            W^{y}_{r}(a^{r}_{x})W^{y}_{r}(O^{r}_{x})
            W^{y}_{r}(b^{r}_{x})=a_{y}O_{y}b_{y},\\\\W^{y}_{r}
            ((a^{r}_{x})^{*^{r}_{x}})=(W^{y}_{r}
            (a^{r}_{x}))^{W^{y}_{r}(*^{r}_{x})}=
            (a_{y})^{*_{y}}\end{array}\end{equation} and
            \begin{equation}\label{Wpropyxx}
            \begin{array}{c}W^{r}_{x}(a_{x})=a^{r}_{x},\\\\
            W^{r}_{x}(a_{x}O_{x}b_{x})=(W^{r}_{x}(a_{x}))W^{r}_{x}
            (O_{x})W^{r}_{x}(b_{x})=a^{r}_{x}O^{r}_{x}b^{r}_{x},
            \\\\W^{r}_{x}(a_{x}^{*_{x}})=(W^{r}_{x}(a_{x}))^{W^{r}_{x}
            (*_{x})}=(a^{r}_{x})^{*^{r}_{x}}.
            \end{array}\end{equation} In these equations $O$ is a
            stand in for the field operations, $\pm,\times,\div.$
            Also $F_{y,x}=W^{y}_{r}W^{r}_{X}$ is an
            isomorphic map from$\bar{C}_{x}$ onto $\bar{C}_{y}.$
            The map $F_{y,x}$ will be referred to as a parallel
            transformation of $\bar{C}_{x}$ to $\bar{C}_{y}$ as it
            defines the notions of "sameness" between $\bar{C}_{x}$
            and $\bar{C}_{y}.$

            Note that $W^{r}_{x}(a_{x})=a^{r}_{x}$ is the same number
            value, $a,$ in $\bar{C}^{r}_{x}$ as $a_{x}$ is in $\bar{C}_{x}$
            and $a_{y}=W^{y}_{r}(a^{r}_{x})$ is the same number value
            in $\bar{C}_{y}$ as $a^{r}_{x}$ is in $\bar{C}_{x}^{r}.$ It follows that
            \begin{equation}\label{ayfyx}a_{y}=F_{y,x}a_{x}\end{equation}
            is the same number value in $\bar{C}_{y}$ as $a_{x}$ is in $\bar{C}_{x}.$

            One still needs to give the explicit correspondence between the number
            values, basic operations, and constants in
            $\bar{C}^{r}_{x}$ and those in $\bar{C}_{x}.$ These
            are given by\begin{equation}\label{opscyxx}
            \begin{array}{c}a^{r}_{x}= ra_{x},\\\\\mbox{} \pm^{r}_{x}
            = \pm_{x},\hspace{1cm}\times^{r}_{x}= \frac{\textstyle\times_{x}}
            {\textstyle r},\\\\ \mbox{}\div^{r}_{x} = r\div_{x},\hspace{1cm}
            (a^{r}_{x})^{*^{r}_{x}}= r(a^{*_{x}}).\end{array}
            \end{equation} The subscripts and superscripts on the number
            values denote their structure membership.

            These equations enable one to express  the elements of
            $\bar{C}^{r}_{x}$ in terms of those in $\bar{C}_{x}$ as
            \begin{equation}\label{CyonCxexpl}\{C,\pm_{x},\frac{\textstyle
            \times_{x}}{\textstyle r},r\div_{x},r(-)^{*_{x}},0_{x},r1_{x}\}.
            \end{equation}Comparison of the elements of $\bar{C}^{r}_{x}$ with
            those of $\bar{C}_{x}$ shows that the number value, $a^{r}_{x}$ in
            $\bar{C}^{r}_{x}$ corresponds to the number value, $ra_{x},$ in
            $\bar{C}_{x},$ where $a_{x}$ is the same number value in $\bar{C}_{x}$
            as $a^{r}_{x}$ is in $\bar{C}_{x}^{r}.$ For example, the identity in
            $\bar{C}^{r}_{x}$ corresponds to the value $r\times_{x}1_{x}=r$ in
            $\bar{C}_{x}$ and multiplication in $\bar{C}^{r}_{x}$ corresponds
            to multiplication divided by $r$ in $\bar{C}_{x}.$ Also the
            relations between the two  structures show that $\bar{C}^{r}_{x}$
            is a scaling of the numbers and operations in $\bar{C}_{x}$ by the
            factor $r.$ The indicated scaling of the operations in
            $\bar{C}_{x}$ compensates for the the fact that the
            number value $a$ (denoted by $a^{r}_{x}$ in $\bar{C}^{r}_{x}$)
            \emph{corresponds} to the  number value $ra_{x}$ in $\bar{C}_{x}.$

            This scaling of the numbers and operations in
            $\bar{C}_{x}$ requires that one drop the condition that
            the elements of the base set $C$ have fixed values,
            independent of the structure containing $C.$ Here, the
            elements in $C$, with one exception, have no fixed value.
            They attain their values only within structures. For example,
            the element (number) in $C$ that has the value $a$ in
            $\bar{C}^{r}_{x},$ has the value $ra$ in $\bar{C}_{x}.$ This
            is equivalent to stating that $a$ in $\bar{C}^{r}_{x}$
            corresponds to $ra$ in $\bar{C}_{x}.$ Also
            the element of $C$ that has the value $a$ in $\bar{C}_{x}$
            is different from the element in $\bar{C}_{x}$ that has the
            same value, $a$ in $\bar{C}^{r}_{x}.$

            The one exception is the element of $C$ that has the value
            $0$. This number has the same value in the structures
            $\bar{C}^{r}_{x}$ for all values of $r.$ In a sense it is
            the "number vacuum". Only for this element can one drop the
            distinction between number and number value.

\begin{figure}[h]\begin{center}
           \resizebox{100pt}{100pt}{\includegraphics[250pt,200pt]
           [500pt,500pt]{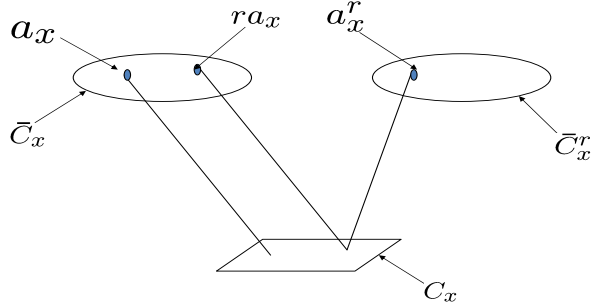}}\end{center}
           \caption{Relations between Elements in the base set $C_{x}$
           and their Numerical Values in  the Structures $\bar{C}_{x}$
           and $\bar{C}^{r}_{x}.$ Here $a^{r}_{x}$ is the same number value in
           $\bar{C}^{r}_{x}$ as $a_{x}$ is in $\bar{C}_{x}.$ As shown by the
           lines they are values for different elements of $C_{x}.$ The
           lines also show that the $C_{x}$ element that has the value $a,$ as
           $a^{r}_{x}$ in $\bar{C}^{r}_{x}$, has the value $ra_{x}$
           in $\bar{C}_{x}.$ Superscripts and subscripts denote structure
            memberships of the values.}\label{NPI1r}\end{figure}

            Some of these relationships are shown explicitly in Figure \ref{NPI1r}.
            It shows explicitly the dependence of the labels or number values of the
            elements of $C_{x}$ on the structure environment.
            These considerations show that the introduction of scaling
            between different complex number structures distinguishes a new relation,
            "correspondence" from that of "same as". The number value
            $ra_{x}$ in $\bar{C}_{x}$ that corresponds to the number value
            $a{r}_{x}$ in $\bar{C}^{r}_{x}$ is different from the number value
            $a_{x}$ in $\bar{C}_{x}$ that is the same value as is $a^{r}_{x}$ in
            $\bar{C}^{r}_{x}.$

            The setup described here collapses to the usual setup with
            one complex number structure at all points if $r_{y,x}=1$
            everywhere. This corresponds to the usual case in which
            the concepts of "correspondence" and "same as" coincide.
            Also  $\bar{C}^{r}_{x} =\bar{C}_{x}$ and
            $W^{y}_{r}=W^{r}_{x}=F_{x,y}=1.$  It follows
            that $\bar{C}_{x}$ for any $x$ is the same as $\bar{C}_{y}$
            for any $y\neq x.$ This is equivalent to saying that $\bar{C}_{x}$
            is independent of $x.$

            The numbers $r_{y,x}$ and $r_{x,y}$ are associated
            with opposite directions of the link between sites $x$
            and $y,$ with $r_{y,x}$ for the direction from $x$
            to $y$ and $r_{x,y}$ for the direction from $y$ to $x.$
            One would like to compare the two numbers.  However, they cannot be
            directly compared because $r_{y,x}$ is a number in $\bar{C}_{x}$
            and $r_{x,y}$ is a number in $\bar{C}_{y}.$

            A comparison can be made between $r_{y,x}$ and
            $(r_{x,y})_{x}=F^{-1}_{y,x}r_{x,y},$ which is the same number in
            $\bar{C}_{x}$ as $r_{x,y}$ is in $\bar{C}_{y}.$ Since
            $(r_{x,y})_{x}$  and $r_{y,x}$ belong to opposite
            directions of the same link, it is
            reasonable to assume that $(r_{x,y})_{x}r_{y,x}=1_{x}$ or
            \begin{equation}\label{invcxycyx}(r_{x,y})_{x}=r_{y,x}^{-1_{x}}.
            \end{equation} The equivalent relation for $\bar{C}_{y}$
            is $(r_{y,x})_{y}=r_{x,y}^{-1_{y}}.$

            So far the description of relations between complex number fields
            has been limited to elements of the fields. However, it can
            be extended to terms and more complicated functions.
            For example, consider the term $(a^{r}_{x})^{m}/
            (b^{r}_{x})^{n}$ in $\bar{C}^{r}_{x}.$  The
            representation of this term in $\bar{C}_{x}$ is given
           by replacing the  factors and  the operations in $\bar{C}^{r}_{x}$
           by their equivalents in $\bar{C}_{x}$ as given in Eq. \ref{opscyxx}.

           The term $( a^{r}_{x})^{m}$ has $m$ factors and $m-1$
           multiplications. These combine to give a factor $r$ so that the
           value $( a^{r}_{x})^{m}$ in $\bar{C}^{r}_{x}$ is the value
           $ra_{x}^{m}$ in $\bar{C}_{x}.$ Combining this with the
           expression $(b^{r}_{x})^{n}= rb_{x}^{n}$
           for the denominator, and a factor of $r$ arising from the
           solidus that represents division, gives the result
           \begin{equation}\label{aynbxm}\frac{(a^{r}_{x})^{m}}
           {(b^{r}_{x})^{n}}\mbox{}^{r}_{x}= r
           \frac{a_{x}^{m}}{b_{x}^{n}}\mbox{}_{x}.\end{equation}
           Since this applies to each term in a power series,
           it applies to the series as a whole. As a result,
           any analytic function $f^{r}_{x}$ on $\bar{C}^{r}_{x}$ corresponds
            to the function, $f_{x},$ in $\bar{C}_{x}$
           multiplied by $r.$ That is \begin{equation}\label{analf}
            f^{r}_{x}(z^{r}_{x})= rf_{x}(z_{x}).\end{equation}

           An example of this is given by the exponential $e^{z}$ as a
           function of the argument, $z.$ The representation of the $\bar{C}^{r}_{x}$
           exponential, $e^{a^{r}_{x}},$  in
           $\bar{C}_{x}$ is given by $re^{a_{x}}.$ This can be
           understood from a power series expansion as
           $$e^{a^{r}_{x}}=\sum_{n}\frac{(a^{r}_{x})^{n}}
           {n^{r}_{x}!}= r\sum_{n}\begin{array}{c}a_{x}^{n}\\\overline{n_{x}!}
           \end{array}.$$This  says that the element of
           $C$ that has value $e^{a}$ in $\bar{C}^{r}_{x}$ has
           value $re^{a}$ in $\bar{C}_{x}.$

           This can be extended to the $\bar{C}_{x}$ representation
           of equations in $\bar{C}^{r}_{x}.$ The above shows that $f^{r}_{x}
           (a^{r}_{x})=b^{r}_{x}$ is the same equation in
           $\bar{C}^{r}_{x}$ as $f_{x}(a_{x})=b_{x}$ is in
           $\bar{C}_{x}.$ This follows from\begin{equation}\label{Eqequal}
           f^{r}_{x}(a^{r}_{x})=b^{r}_{x}\Leftrightarrow
           rf_{x}(a_{x})=rb_{x}\Leftrightarrow f_{x}(a_{x})=b_{x}.
           \end{equation}  This result is important because it  shows
           that the local representations,  in $\bar{C}_{x},$ of
           equations in $\bar{C}_{y}$  are the same
           equations as those in $\bar{C}_{x}$ that are obtained by
           parallel transformation of equations in $\bar{C}_{y}.$

           As was noted before, the  relations between the basic
           operations in $\bar{C}^{r}_{x}$ and those in $\bar{C}_{x},$
           as seen in Eq. \ref{opscyxx}, must be such that $\bar{C}_{x}$
           satisfies the  complex number axioms \cite{Adamson} if
           and only if $\bar{C}^{r}_{x}$ satisfies the axioms. The
           validity of this requirement is a consequence of the fact
           that all the complex number axioms are equations. As was
           seen above, equations are valid in $\bar{C}^{r}_{x}$ if
           and only if their corresponding representations in
           $\bar{C}_{x}$ are valid.

            A couple of examples of proofs for individual axioms are
            sufficient, as proofs to the other axioms are similar.
            For the axiom, $a\times a^{-1}=1,$ one has the following
            equivalences: $$\begin{array}{l}a^{r}_{x}\times^{r}_{x}
            (1^{r}_{x}\div^{r}_{x}a^{r}_{x})=
            1^{r}_{x}\Leftrightarrow(ra_{x})\times^{r}_{x}(r1_{x})
            \div^{r}_{x}(ra_{x})=(r1_{x})\\\\\Leftrightarrow (ra_{x})
            (\frac{\textstyle \times_{x}}{\textstyle r}\mbox{}_{x})(r1_{x})
            (\frac{\textstyle r}{\textstyle ra_{x}}\mbox{}_{x})=r1_{x}
            \Leftrightarrow a_{x}\times_{x} (1_{x}\div_{x}a_{x})=1_{x}.
            \end{array}$$ Here Eq.\ref{opscyxx} was used to obtain these
            equivalences. For algebraic closure one can show that $a_{x}$
            is the solution of a polynomial equation $P_{x}(z)=0$
            in $\bar{C}_{x}$ if and only if $a^{r}_{x}$ is a solution
            of the corresponding polynomial equation $P^{r}_{x}
            (z^{r}_{x})=0$ in $\bar{C}^{r}_{x}.$

            The involution axiom $(a^{*})^{*}=a$ is another example.
            From Eq. \ref{opscyxx} one has \begin{equation}
            \label{acxxstar}((a^{r}_{x})^{*^{r}_{x}}
            )^{*^{r}_{x}}=((ra_{x})^{*^{r}_{x}})^{*^{r}_{x}}
            =(r(a_{x}^{*_{x}}))^{*^{r}_{x}}=r(a_{x})^{*_{x}*_{x}}.
            \end{equation} From this one obtains the equivalences
            $$(a^{r}_{x})^{*^{r}_{x}*^{r}_{x}}=a^{r}_{x}\Leftrightarrow
            (r(a_{x}^{*_{x}}))^{*^{r}_{x}}=a^{r}_{x}\Leftrightarrow
            r(a_{x})^{*_{x}*_{x}}=ra_{x}\Leftrightarrow
            (a_{x}^{*_{x}})^{*_{x}}=a_{x}.$$

           \section{Gauge Fields}\label{GF}
            The association of different complex number structures
            to space time points can be represented as a
            field, $\mathfrak{F},$ over space time of complex number structures.
            The association is given by $\mathfrak{F}:x\rightarrow\bar{C}_{x}.$

            From the viewpoint of an observer at $x$ for whom the
            elements of $\bar{C}_{x}$ are the complex numbers, there
            is a local representation, $\mathfrak{F}_{x},$ of the
            field, $\mathfrak{F}.$ This consists of the set of local representations
            of $\bar{C}_{y}$  on $\bar{C}_{x}$ for all space time points $y$,
            not just those that are neighbors of $x.$ For points distant from
            $x$ the superscript $r_{y,x}$ in $\bar{C}^{r_{y,x}}_{x}$ is
            replaced by an integral over paths from $x$ to $y.$ These are
            discussed in the next subsection. Here the map, $W^{r}_{x},$ is a
            connection, or element of the tangent space on $\mathfrak{F}_{x}.$

            The $W^{r_{y,x}}_{x}$ are elements of the gauge group $GL(1,R).$
            Gauge fields enter through the representation of $W^{r}_{x}$
            as \begin{equation}\label{vecAx}
            W^{r}_{x}=e^{\vec{A}(x)\cdot\hat{\nu}dx}=e^{A_{\mu}(x)
            dx^{\mu}}.\end{equation}(Sum over $\mu$ implied.) Here
            $y=x+\hat{\nu}dx$ is a neighbor point of $x$ and
            $\vec{A}(x)$ is a real valued  gauge field with four space
            time components $A_{\mu}(x).$ These components are real
            numbers in $\bar{C}_{x}$ and are associated with
            the link from $x$ to $y.$

            Since  $W^{r}_{x}a_{x}=r_{y,x}a_{x},$ Eqs. \ref{Wpropyxx} and
            \ref{opscyxx}, one has\begin{equation}\label{Wcaxe}W^{r}_{x}a_{x}=
            e^{\vec{A}(x)\cdot\hat{\nu}dx}a_{x}.\end{equation} This gives
            \begin{equation}\label{cnuvecA}r_{y,x}=e^{\vec{A}(x)
            \cdot\hat{\nu}dx},\end{equation} which is a repetition of
            Eq. \ref{rvecA}.

            One can use the relation between $r_{y,x}$ and $(r_{x,y})_{x}$
           to obtain a corresponding relation for the  gauge
           field for $(r_{x,y})_{x}$. From $r_{y,x}(r_{x,y})_{x}=
           1_{x}$ one has\begin{equation}\label{cxyx}(r_{x,y})_{x}=
           1_{x}/r_{y,x}=e^{-\vec{A}(x)\cdot\hat{\nu}dx}.
           \end{equation}this shows that if $\vec{A}(x)$ is associated with the
           link from $x$ to $y$, then $-\vec{A}(x)$ is associated with the same
           link in the opposite direction.  In either case the components
           of $\vec{A}(x)$ are real number values in $\bar{C}_{x}.$

           To first order in small quantities, $W^{r}_{x}$, or $r_{y,x},$
           ($W^{r}_{x}$ and $r_{y,x}$ are used interchangeably here)
           can be expressed by \begin{equation}\label{cyx1stotd}W^{r}_{x}
           =r_{y,x}=1+\vec{A}(x)\cdot \hat{\nu}dx.\end{equation} This
           shows that for $y$ a neighbor point of $x,$ $r_{y,x}$
           corresponds to a scale change factor,
           $1+\vec{A}(x)\cdot\hat{\nu}dx,$ in going from
           $\bar{C}^{r}_{x}$ to $\bar{C}_{x}.$

           \subsection{Representations of Numbers
           at Distant Points.}\label{RNVDP}
           So far the description is limited to  $x$ representations
           of number structures at neighbor points $y.$ This needs
           to be extended to the case where $y$ is arbitrary. The fact
           that number values associated with different points belong to
           different complex number structures needs to be taken into
           account.

           To begin, consider a two step path $x\rightarrow y\rightarrow z$
           where $y=x+\hat{\nu}_{x}\Delta_{x}$ and $z=y+\hat{\nu}_{y}
           \Delta_{y}.$ Subscripts will be left off of the small quantity,
           $\Delta,$ because it is the same  number at $y$ as at $x$
           ($\Delta_{y}=F_{y,x}(\Delta_{x})=\Delta$). Let $a_{z}$ denote
           a number  in $\bar{C}_{z}.$ The goal is to find the
           number in $\bar{C}_{x}$ that corresponds to $a_{z}.$

           Here and from now on, number values will often be referred to as
           numbers. The reason is that as it will be clear from context
           whether one is referring to elements of a base set or the
           values the elements take in a given structure. In cases where
           it is not clear, number value will be used.

           From Eqs. \ref{opscyxx} and \ref{Rppropyxx} one sees that
           the number  in $\bar{C}_{y}$  that corresponds to  $a_{z}$
           is  $(W_{r_{z,y}}^{z})^{-1}(a_{z})$. It follows that the
           number in $\bar{C}_{x}$ that corresponds to  $a_{z}$ is
           \begin{equation}\label{RpyxRpzy}\begin{array}{l}
           (W_{r_{y,x}}^{y})^{-1}((W_{r_{z,y}}^{z})^{-1}(a_{z}))=
           (W_{r_{y,x}}^{y})^{-1}(r_{z,y}\times_{y}a_{y})\\\\\hspace{1cm}=
           W^{r_{y,x}}_{x}([r_{z,y}]_{x}a_{x})=r_{y,x}[r_{z,y}]_{x}
           a_{x}.\end{array}\end{equation} Here $[r_{z,y}]_{x}=
           F_{x,y}(r_{z,y})$ is the same number in $\bar{C}_{x}$
           as $r_{z,y}$ is in $\bar{C}_{y}.$

           The  gauge field expression for $r_{y,x}[r_{z,y}]_{x}
           a_{x}$ is\begin{equation}\label{RpRpvecA}r_{y,x}[r_{z,y}]_{x}
           a_{x}=e^{\vec{A}(y)_{x} \cdot\hat{\nu}_{y}\Delta +\vec{A}(x)
           \cdot\hat{\nu}_{x}\Delta}a_{x}.
           \end{equation} Here $a_{x}=F_{x,z}a_{z}$ is the same number in
           $\bar{C}_{x}$ as $a_{z}$ is in $\bar{C}_{z}.$ The commutativity
           of $\vec{A}(x)$ with $\vec{A}(y)_{x}$ is used here along with
           the observation that\begin{equation}\label{evec}[r_{z,y}]_{x}=
           [e^{\vec{A}(y)\cdot\hat{\nu}_{y}\Delta}]_{x}=e^{\vec{A}(y)_{x}
           \cdot\hat{\nu}_{y}\Delta}.\end{equation}  The components,
           $A_{\mu}(y)_{x},$ of $\vec{A}(y)_{x}$ are the same
           numbers in $\bar{C}_{x}$ as the $A_{\mu}(y)$ are in $\bar{C}_{y}.$

           The notation used here is that for
           functions, such as $A_{\mu}(y),$ the subscript $x,$ as in
           $A_{\mu}(y)_{x},$ denotes the same number value in $\bar{C}_{x}$
           as $A_{\mu}(y)$ is in $\bar{C}_{y}.$ For number values with
           subscripts, such as $r_{z,y},$ square brackets and a subscript,
           such as $[r_{z,y}]_{x}$ are used to denote structure membership.

           Extension of the two step result to an $n$ step path $P$ where
           $P(0)=x_{0}=x,P(n)=x_{n}=y,P(j)=x_{j}$ and $x_{j+1}=x_{j}+\hat{\nu}_{j}
           \Delta$ for $0\leq j\leq n$ gives \begin{equation}\label{RepxfycP}
           W^{x}_{y}(a_{y})=r^{P}_{y,x}a_{x}\end{equation} where
           \begin{equation}\label{cPyx}r^{P}_{y,x}=\prod_{j=0}^{n-1}
           [r_{x_{j+1},x_{j}}]_{x}.\end{equation}Here $W^{x}_{y}(a_{y})$
           denotes the representation of $a_{y}$ in $\bar{C}_{x}.$ The gauge field
           expression for $r^{P}_{y,x}$ is\begin{equation}\label{GgecPyx}
           r^{P}_{y,x}=\exp(\sum_{j=0}^{n-1}[\vec{A}(x_{j})
           \cdot\hat{\nu}_{j}\Delta_{x_{j}}]_{x}).\end{equation}

           For a continuous path $P$ from $x$ to $y$ Eq.
           \ref{GgecPyx} becomes\begin{equation}\label{cPyxAint}
           r^{P}_{y,x}=\exp\{\int_{0}^{1}\vec{A}(P(s))_{x}\cdot
           [\frac{dP(s)} {d s}]_{x}ds\}.\end{equation} The path
           is parameterized by a variable $s,$ $0\leq s\leq 1,$
           where $P(0)=x$ and $P(1)=y.$ An ordering of the integrand variables
           is not needed because the $\vec{A}(y)_{x}$ commute for different $y.$

           The subscript $x$ in the integrand  and on $\vec{A}(P(s))_{x}$
           indicate that the integral is defined in $\bar{C}_{x}$ and that
           the components, $A_{\mu}(P(s))_{x}$, which are numbers in
           $\bar{C}_{x},$ are the same numbers in $\bar{C}_{x}$ as
           the numbers $A_{\mu}(P(s))$ are in $\bar{C}_{P(s)}.$ This
           is expressed by\begin{equation}\label{APsxAps}
           A_{\mu}(P(s))_{x}=F_{x,P(s)}A_{\mu}(P(s)).\end{equation}
           The derivative components, $[dP_{\mu}(s)/ds]_{x},$ which
           are the same numbers in $\bar{C}_{x}$
           as $dP_{\mu}(s)/ds,$ are in $\bar{C}_{P(s)}$ are
           related to $dP_{\mu}(s)/ds$ by\begin{equation}
           \label{dPmusx}[\frac{dP_{\mu}(s)}{ds}]_{x}=F_{x,P(s)}
           \frac{dP_{\mu}(s)} {ds}.\end{equation}

           $r^{P}_{y,x}$ can be expressed as the exponential of
           a line integral along the path $P$ as in\begin{equation}\label{LIcPyx}
           r^{P}_{y,x}=\exp(\int_{P}\vec{A}(z)_{x}\vec{dz}).
           \end{equation}The subscript $x$ indicates that the integral
           is defined in $\bar{C}_{x}.$

           \subsection{Derivatives and Integrals over Space Time}
           \label{STI}

           The dependence of $\bar{C}_{x}$ on $x$ has an effect on
           derivatives and integrals of functions over space time.
           To see this let $\Phi(x)$ be a function over space time
           such that for each $x,$ $\Phi(x)$ is a number in
           $\bar{C}_{x}.$ As was noted in the introduction for
           Eq. \ref{dmux}, the problems of the the usual
           derivative,\begin{equation}\label{usualpar}\partial_{\mu,x}\Phi=
           \frac{\Phi(x+dx^{\mu})-\Phi(x)}{\partial{x^{\mu}}}
           \end{equation} can be avoided by
           defining a local representation of $\bar{C}_{x+dx^{\mu}}$
           on $\bar{C}_{x}$ as in Eq. \ref{CyonCxexpl}. This enables
           the derivative to be expressed entirely within $\bar{C}_{x}$
           as, Eq. \ref{Dmuxp},
           \begin{equation}\label{DPhi}D_{\mu,x}\Phi=\frac{r_{x+dx^{\mu}
           ,x}\Phi(x+dx^{\mu})_{x}-\Phi(x)}{\partial{x}^{\mu}}
           \end{equation} where\begin{equation}\label{Phimuxx}
           \Phi(x+dx^{\mu})_{x}=F_{x+dx^{\mu},x}^{-1}\Phi(x+dx^{\mu})
           \end{equation} is the same number value in $\bar{C}_{x}$ as
           $\Phi(x+dx^{\mu})$ is in $\bar{C}_{x+dx^{\mu}}.$ The term,
           $r_{x+dx^{\mu},x}\Phi(x+dx^{\mu})_{x},$
           in Eq. \ref{DPhi}, denotes the number in $\bar{C}_{x}$
           that corresponds to the number $\Phi(x+dx^{\mu})$ in
           $\bar{C}_{x+dx^{\mu}}.$

           Expressing $r_{x+dx^{\mu},x}$ in terms of
           gauge potentials and expanding to first order in
           $dx^{\mu}$ gives\begin{equation}\label{DmuxPa}
           D_{\mu,x}\Phi=(\partial^{\prime}_{\mu,x}+A_{\mu}(x))
           \Phi(x)\end{equation} where\begin{equation}\label{parprim}
           \partial^{\prime}_{\mu,x}\Phi=\frac{\Phi(x+dx^{\mu})_{x}-
           \Phi(x)}{\partial{x}^{\mu}}\end{equation} is a repetition
           of Eq. \ref{dmuxp}.

           The presence of  the gauge field, $\vec{A}(x),$  is a
           consequence of the freedom to choose complex number
           structures, one for each $x.$ If $\vec{A}(x)=0$ everywhere, then
           the complex number structures would all be the same and
           they can be collapsed into one structure. In this case
           $F_{y,x}=1$ and\begin{equation}\label{Dpartial}
           D_{\mu,x}\Phi=\partial^{\prime}_{\mu,x}\Phi=\partial_{\mu,x}
           \Phi.\end{equation}

           Similar considerations hold for space time integrals. The
           usual expression $\int \Phi(x)d^{4}x.$ is not defined because
           it corresponds to adding together values of $\Phi(x)$
           in different $\bar{C}_{x}.$
           One way to handle this is to choose a point $x$ and the
           associated $\bar{C}_{x}$ and map all the values of
           $\Phi(y)$ to their representations in
           $\bar{C}_{x}.$ The integral is then defined on
           $\bar{C}_{x}.$ This is done by replacing $\Phi(y)$ by its
           local representative in $\bar{C}_{x},$ which is
           \begin{equation}\label{Repxfy}W^{x}_{y}(\Phi(y))=r^{P}_{y,x}
           F^{-1}_{y,x}(\Phi(y))=r^{P}_{y,x}\Phi(y)_{x}.\end{equation}
           Here $\Phi(y)_{x}$ is the $F$ same number in $\bar{C}_{x}$
           as $\Phi(y)$ is in $\bar{C}_{y}$ and $P$ is a path from $x$
           to $y.$ $c^{P}_{y,x}$ is given by Eq. \ref{LIcPyx}.

           Putting this together gives\begin{equation}\label{intx}
           \int_{x} \Phi=\int r^{P}_{y,x}\Phi(y)_{x}d^{4}y=\int
           \exp(\int_{P}\vec{A}(z)_{x}\vec{dz})\Phi(y)_{x}d^{4}y.\end{equation}
            This expression has the disadvantage that it depends on
            paths from all space time points to $x$. One might be
            able to handle this by carrying out some type of Feynman
            path integral.  This can be avoided if one assumes
            that the $r^{P}_{y,x}$ are integrable, or path
            independent.  In this case Eq. \ref{intx} becomes
            \begin{equation}\label{intxfPI}\int_{x} \Phi=\int
            r_{y,x}\Phi(y)_{x}d^{4}y=\int (e^{\int^{y}_{x} \vec{A}})\Phi(y)_{x}
            d^{4}y.\end{equation}Here the line integral of $\vec{A}$
            is independent of the path choice. However, it remains to
            be seen if the assumption of integrability is valid or
            not.

            There remains the dependence on $x$ indicated by the
            subscript $x$ on $\int_{x} \Phi.$ It is suspected
            that the results are independent of $x.$ This is based
            on the observation that all the results obtained so
            far depend on the relation between systems at $x$ and $y$
            and not on the location of $x.$  However, this remains
            to be investigated.

              \section{Hilbert Spaces}\label{HSSF}

            So far, complex number structures have been considered
            by themselves.  However, there are many types of mathematical
            systems that are based on the real or complex numbers as
            underlying scalar fields. Vector spaces based on complex numbers
            are examples of this type of mathematical system.  For these
            spaces, it would be  expected that the differences between
            complex number systems at different space time points would
            have an effect on vectors and vector spaces assigned to
            different points.

            Here this is seen to be the case for Hilbert spaces as
            examples of vector spaces. Let $\bar{H}_{x},\bar{C}_{x}$
            be an $n$ dimensional Hilbert space and a complex number
            field at point $x.$ As a structure
           \begin{equation}\label{barHx}\bar{H}_{x}=\{H_{x},+_{x},
           -_{x},\cdot_{x}, \langle -,-\rangle_{x},\psi_{x}\}
           \end{equation} where $H_{x}$ is a base set,\footnote{Here
           the term "vector" will be used to denote the values of
           the elements in the base set $H_{x}.$} $+_{x}$ and
           $-_{x}$ denote linear addition
           and subtraction, $\cdot_{x}$ denotes multiplication of a
           vector by a scalar in $\bar{C}_{x},$ and $(-,-)_{x}$
           denotes the scalar product with values in $\bar{C}_{x}.$
           $\psi_{x}$ denotes an arbitrary vector in $\bar{H}_{x}.$

           The basic operations  shown in Eq. \ref{barHx} must satisfy
           the axioms for a Hilbert space. These describe a complex
           inner product vector space that is complete in the
           norm \cite{Kadison1}.

           The freedom  of basis choices for Hilbert
           spaces at different space time points \cite{Yang}
           can be expressed here by representing the parallel
           transform, $U_{y,x},$ from $\bar{H}_{x}$ to $\bar{H}_{y}$
           as a product of two factors, as in \begin{equation}
           \label{HyMfrkVHx}\bar{H}_{y}=U_{y,x}\bar{H}_{x}=
           \mathfrak{V}_{y,x}V_{y,x}\bar{H}_{x}.\end{equation}

           The reason that the unitary operator, $U_{y,x},$ is  a
           product of two factors is that $U_{y,x}$
           cannot be represented as a unitary matrix
           of complex number entries. If $U_{y,x}=\{a_{i,j,x}\}$ where the
           $a_{i,j,x}$ are numbers in $\bar{C}_{x},$ then for any vector
           $\psi_{x}$ in $\bar{H}_{x},$ $U_{y,x}\psi_{x}=\sum_{i,j}
           |i\rangle_{x}a_{i,j,x}\langle j,\psi_{x}\rangle_{x}$ is a vector
           in $\bar{H}_{x}.$ It is not a vector in $\bar{H}_{y}.$
           This is the case even if $\bar{C}_{x}$ is replaced by
           $\bar{C},$ which is common to both $\bar{H}_{x}$ and
           $\bar{H}_{y}.$

           The unitary operator $V_{y,x}$ maps  a set of basis
           vectors to a transformed basis
           in $\bar{H}_{x}$ that is the $\bar{H}_{x}$
           representation of a basis in $\bar{H}_{y}.$ If
           $|j\rangle_{x}$ is a basis vector in $\bar{H}_{x}$ then
           $V_{y,x}|j\rangle_{x}$ is the transformed vector that is
           the representation of $|j\rangle_{y}$ in
           $\bar{H}_{x}.$ $\mathfrak{V}_{y,x}$ maps the transformed
           basis to the corresponding basis in $\bar{H}_{y}$ that is
           the same as the original basis in $\bar{H}_{x}.$
           That is, $|j\rangle_{y}=U_{y,x}|j\rangle_{x}.$

           Eq. \ref{HyMfrkVHx} describes parallel transformations
           from $\bar{H}_{x}$ to $\bar{H}_{y}$. However it does not
           include the corresponding changes in going from $\bar{C}_{x}$
           to $\bar{C}_{y}.$ These are taken into account by considering
           the overall transformation $\bar{H}_{x},\bar{C}_{x}$ to
           $\bar{H}_{y},\bar{C}_{y}$ as a three step process.
           The first step takes $\bar{H}_{x}$ to $V_{y,x}
           \bar{H}_{x}.$ $\bar{C}_{x}$ remains unchanged.

            The second step takes $V_{y,x}\bar{H}_{x}$
            to $\mathfrak{V}^{r}_{x}V_{y,x}\bar{H}_{x}=
            \bar{H}^{r}_{x}$ and $\bar{C}_{x}$ to
            $W^{r}_{x}\bar{C}_{x}=\bar{C}^{r}_{x}$ and the
            third step takes $\bar{H}^{r}_{x}$ and $\bar{C}^{r}_{x}$
            to $\bar{H}_{y}=\mathfrak{V}_{y,x}\bar{H}^{r}_{x}$
            and $\bar{C}_{y}=W^{y}_{r}\bar{C}^{r}_{x}.$ Here
            $\bar{H}^{r}_{x}$ and $\bar{C}^{r}_{x}$ are the
            respective local representations of $\bar{H}_{y}$ and
            $\bar{C}_{y}$ on $\bar{H}_{x}$ and $\bar{C}_{x}.$
           A summary of the three steps is given by\begin{equation}\label{3stepcyx}
           \begin{array}{c}\bar{H}_{x}\\\bar{C}_{x}\end{array}\rightarrow
           \begin{array}{c}V_{y,x}\bar{H}_{x}\\\bar{C}_{x}\end{array}
           \rightarrow\begin{array}{c}\mathfrak{V}^{r}_{x}V_{y,x}\bar{H}_{x}
           \\W^{r}_{x}\bar{C}_{x}\end{array}=\begin{array}{c}\bar{H}^{r}_{x}
           \\\bar{C}^{r}_{x}\end{array}\rightarrow
           \begin{array}{c}\mathfrak{V}^{y}_{r}\bar{H}^{r}_{x}\\W^{y}_{r}
           \bar{C}^{r}_{x}\end{array}=\begin{array}{c}\mathfrak{V}_{y,x}V_{y,x}
           \bar{H}_{x}\\F_{y,x}\bar{C}_{x}\end{array}=
           \begin{array}{c}\bar{H}_{y}\\\bar{C}_{y}\end{array}.
           \end{equation}Here $W^{r}_{x}$
           and $W^{y}_{r}$ are given by Eqs. \ref{FxyWx}-\ref{Wpropyxx}.

             $\bar{H}^{r}_{x}$ has the structure representation
            \begin{equation}\label{Hcyxx}\bar{H}^{r}_{x}=\{H_{x},
           \pm^{r}_{x},\cdot^{r}_{x},\langle-,-\rangle^{r}_{x},
           \psi^{r}_{x}\}.\end{equation}$H_{x}$ is the same base set
           as that in $\bar{H}_{x},$ $\pm^{r}_{x}$ denotes
           addition and subtraction of vectors, $\cdot^{r}_{x}$ and
           $\langle -,-\rangle^{r}_{x}$ denote scalar vector
           multiplication and scalar product, and $\psi^{r}_{x}$ denotes
           an arbitrary vector in $\bar{H}^{r}_{x}.$ Also
           $\bar{C}^{r}_{x}$ is given by Eq. \ref{Ccxatx}, and, as before,
           $r=r_{y,x}$.

           The map $\mathfrak{V}^{r}_{x}$ is  defined by\begin{equation}
           \label{MFRKVcxprop}\begin{array}{c}\mathfrak{V}^{r}_{x}V_{y,x}
           \psi_{x}=\psi^{r}_{x}\\\\\mathfrak{V}^{r}_{x}
           (V_{y,x}\psi_{x}\pm_{x}V_{y,x}\phi_{x})=(\mathfrak{V}^{r}_{x}
           V_{y,x}\psi_{x})\pm^{r}_{x}(\mathfrak{V}^{r}_{x}V_{y,x}
           \phi_{x})=\psi^{r}_{x}\pm^{r}_{x}\phi^{r}_{x}\\\\\mathfrak{V}^{r}_{x}(
           \alpha_{x}\cdot_{x}V_{y,x}\psi_{x})=(W^{r}_{x}\alpha_{x})\cdot^{r}_{x}
           \mathfrak{V}^{r}_{x}V_{y,x}\psi_{x}=\alpha^{r}_{x}
           \cdot^{r}_{x}\psi^{r}_{x}\\\\\mathfrak{V}^{r}_{x}
           \langle V_{y,x}\psi_{x},V_{y,x}\phi_{x}\rangle_{x}=
           \langle\mathfrak{V}^{r}_{x}V_{y,x}\psi_{x},\mathfrak{V}^{r}_{x}
           V_{y,x}\phi_{x}\rangle_{x}^{r}=\langle\psi^{r}_{x},\phi^{r}_{x}
           \rangle^{r}_{x}.\end{array}\end{equation}

           As was the case of the complex number structures, the
           representation of $\bar{H}_{y},\bar{C}_{y}$ on
           $\bar{H}_{x},\bar{C}_{x}$  is obtained by giving the
           definitions of the vectors and basic operations in
           $\bar{H}^{r}_{x}$ in terms of the vectors and basic
           operations in $\bar{H}_{x}.$  The resulting representation
           of $\bar{H}^{r}_{x}$  must also be such that
           $\bar{H}^{r}_{x}$ satisfies the Hilbert space axioms
           if and only if $\bar{H}_{x}$ satisfies the axioms.

           The definitions are made specific by first giving  a specific
           representation of the vector $\psi^{r}_{x}$ in $\bar{H}_{x}.$
           The development so far suggests that both $V_{y,x}$ and the
           dependence of the complex numbers on $x$ should be included.
           This can be achieved by the specific correspondence
           \begin{equation}\label{psicxMFK} (\psi^{r}_{x})_{x}=\mathfrak{V}^{r}_{x}
           V_{y,x}\psi_{x}= r_{y,x}V_{y,x}\psi_{x}=e^{\vec{A}(x)\cdot
           \hat{\nu}dx}V_{y,x}\psi_{x}.\end{equation}The parentheses
           around $\psi^{r}_{x}$ and the subscript $x$ indicate that
           $(\psi^{r}_{x})_{x}$ is the vector in $\bar{H}_{x}$ that
           corresponds to $\psi^{r}_{x}$ in $\bar{H}^{r}_{x}.$\footnote{\label{Cn}
           Support for Eq. \ref{psicxMFK} comes from the Hilbert
           space complex number equivalence $\bar{H}\simeq\bar{C}^{n}$
           \cite{Kadison}. Here $\bar{H}_{x}\simeq(\bar{C}_{x})^{n}$
           and $\bar{H}^{r}_{x}\simeq(\bar{C}^{r}_{x})^{n}.$
           As a simple case let $V_{y,x}=1.$ Then vectors in
            $(\bar{C}^{r}_{x})^{n}$ consist of $n-tuples$
            $(a_{1})^{r}_{x}\cdots (a_{n})^{r}_{x}$ of complex
            numbers in $\bar{C}^{r}_{x}.$ Since $(a_{j})^{r}_{x}$
            corresponds to the number $r(a_{j})_{x}$ in $\bar{C}_{x}$
            for $j=1,\cdots,n$, $\bar{a}^{r}_{x}=(a_{1})^{r}_{x}\cdots
            (a_{n})^{r}_{x}$ corresponds to the vector $r[(a_{1})_{x}\cdots
            (a_{n})_{x}]=r\bar{a}_{x}$ in $(\bar{C}_{x})^{n}.$
            Here $ (a_{j})_{x}$ is the same number in $\bar{C}_{x}$
            as $(a_{j})^{r}_{x}$ is in $\bar{C}^{r}_{x}.$
            The scalar product of two vectors, $\bar{a}^{r}_{x},$
            $\bar{b}^{r}_{x}$ is a number in $\bar{C}^{r}_{x},$ given
            by $$\langle \bar{a}^{r}_{x},\bar{b}^{r}_{x}\rangle^{r}_{x}=
            \sum_{j}((a_{j})^{r}_{x})^{*^{r}_{x}}\times^{r}_{x}
            (b_{j})^{r}_{x}.$$ This corresponds to the number
            $$r\sum_{j}((a_{j})_{x})^{*_{x}}\times_{x}
            (b_{j})_{x}=r\langle\bar{a}_{x},\bar{b}_{x}\rangle_{x}$$
            in $\bar{C}_{x}.$ The correspondence reflects the fact
            that $\langle\bar{a}^{r}_{x},\bar{b}^{r}_{x}\rangle^{r}_{x}$
            is the same number in $\bar{C}^{r}_{x}$ as $\langle
            \bar{a}_{x},\bar{b}_{x}\rangle_{x}$ is in $\bar{C}_{x}.$}
            Alternatively one can say that $(\psi^{r}_{x})_{x}$ is the
            local representation of $\psi_{y}$ on $\bar{H}_{x}$ where
            $\psi_{y}$ is the same state in $\bar{H}_{y}$ as $\psi_{x}$
            is in $\bar{H}_{x}.$

           Reference to Fig. \ref{NPI1r}, which also applies to other
           mathematical systems, such as Hilbert spaces, is useful here.
           It shows that the element of the base set $H_{x},$ that is
           the vector $\psi^{r}_{x}$ in $\bar{H}^{r}_{x},$ is the vector
           $r_{y,x}V_{y,x}\psi_{x}$ in $\bar{H}_{x}.$ Also the
           element of $H_{x}$ that is the vector $\psi^{r}_{x}$ in
           $\bar{H}^{r}_{x}$ is different from the element that is
           the same vector $\psi_{x}$ in $\bar{H}_{x}.$

           The requirement that $rV_{y,x}\psi_{x}$ in $\bar{H}_{x}$,
           corresponds to the vector $\psi^{r}_{x}$ in $\bar{H}^{r}_{x},$ results
           in compensatory changes in basic operations in $\bar{H}^{r}_{x},$
           expressed in terms of operations in $\bar{H}_{x}.$ The
           changes must be such that the structure, $\bar{H}^{r}_{x},$
           satisfies the Hilbert space axioms if and only if $\bar{H}_{x}$ does.

           Eq. \ref{psicxMFK} can be used to determine the relations between
           the Hilbert space operations in $\bar{H}_{x}$ and those in
           $\bar{H}^{r}_{x}.$ The representation, in $\bar{H}_{x}$  of
           the linear superposition operation in $\bar{H}^{r}_{x},$
           is given
           by\begin{equation}\label{pmcyxxx}(\pm^{r}_{x})_{x}
           =\pm_{x}.\end{equation} This follows from the equivalences
           $$\begin{array}{l}\psi^{r}_{x}\pm^{r}_{x}\phi^{r}_{x}=\theta^{r}_{x}
           \Leftrightarrow r_{y,x}V_{y,x}\psi_{x}(\pm^{r}_{x})_{x}
           r_{y,x}V_{y,x}\phi_{x}
           =r_{y,x}V_{y,x}\theta_{x}\\\hspace{1cm}\Leftrightarrow
           \psi_{x}\pm_{x}\phi_{x}=\theta_{x}.\end{array}$$

            For scalar vector multiplication one can use the equation,
            $\psi_{x}=1_{x}\cdot_{x}\psi_{x}$ in $\bar{H}_{x}$ to
            determine the relations.  The equivalences
           $$\begin{array}{l}\psi^{r}_{x}=1^{r}_{x}
           \cdot^{r}_{x}\psi^{r}_{x}\Leftrightarrow r
           V_{y,x}\psi_{x}=(r1_{x})(\cdot^{r}_{x})_{x}(r
           V_{y,x}\psi_{x})\\\hspace{1cm}\Leftrightarrow
           \psi_{x}=1_{x}\cdot_{x}\psi_{x}\end{array}$$ require
           that $r(\cdot^{r}_{x})_{x} =\cdot_{x},$ or
           \begin{equation}\label{cdotcx}(\cdot^{r}_{x})_{x}=
           \frac{\cdot_{x}}{r}.\end{equation}

           For scalar products one requires  $\langle\psi_{x}^{r}
           ,\phi_{x}^{r}\rangle^{r}_{x}$ to be the same
           number in $\bar{C}^{r}_{x}$ as $\langle\psi_{x},
           \phi_{x}\rangle_{x}$ is in $\bar{C}_{x}.$  This is based
           on the fact that $\psi_{x}^{r}$ is the same vector in
           $\bar{H}^{r}_{x}$ as $\psi_{x}$ is in $\bar{H}_{x}.$

           This requirement (footnote on page \pageref{Cn}) is
           expressed by the equation equivalences,
           \begin{equation}\label{psiyxphiyx}\begin{array}{l}\langle
           \psi_{x}^{r},\phi_{x}^{r}\rangle^{r}_{x}=d^{r}_{x}\Leftrightarrow
           (\langle\psi_{x}^{r},\phi_{x}^{r}\rangle^{r}_{x})_{x}=rd_{x}
           \\\hspace{1cm}\Leftrightarrow\langle\psi_{x},\phi_{x}\rangle_{x}
           =d_{x}.\end{array}\end{equation} Here $(\langle \psi_{x}^{r},
          \phi_{x}^{r}\rangle^{r}_{x})_{x}$ is the representation, in $\bar{C}_{x},$
          of the number value, $\langle\psi_{x}^{r},
          \phi_{x}^{r}\rangle^{r}_{x},$ in $\bar{C}^{r}_{x}.$

          It remains to determine the relationship between $(\langle
          \psi_{x}^{r},\phi_{x}^{r}\rangle^{r}_{x})_{x}$ and
          $\langle(\psi_{x}^{r})_{x},(\phi_{x}^{r})_{x}\rangle_{x}.$
          From Eq. \ref{psicxMFK} one has\begin{equation}\label{psixrxx}
          \langle(\psi_{x}^{r})_{x},(\phi_{x}^{r})_{x}\rangle_{x}=
          \langle rV_{y,x}\psi_{x},rV_{y,x}\phi_{x}\rangle_{x}=r^{2}
          \langle\psi_{x},\phi_{x}\rangle_{x}.\end{equation} Use of
          Eq. \ref{psiyxphiyx} gives\begin{equation}\label{xrxx}
          \langle(\psi_{x}^{r})_{x},(\phi_{x}^{r})_{x}\rangle_{x}
          =r(\langle\psi_{x}^{r},\phi_{x}^{r}\rangle^{r}_{x})_{x}.
          \end{equation}

           These relations do not  conflict with the usual properties
           of scalar products. For example, the $\bar{C}^{r}_{x}$
           norm of a vector is the same number as is the
           corresponding norm in $\bar{C}_{x}.$  This can be
           seen from Eq. \ref{psiyxphiyx} which gives $$
           \langle\psi^{r}_{x},\psi^{r}_{x}\rangle^{r}_{x}=1^{r}_{x}
           \Leftrightarrow\langle \psi_{x},\psi_{x}\rangle_{x}=1_{x}.$$
           Norm preservation occurs because $1^{r}_{x}=r1_{x}$
           is the multiplicative identity in $\bar{C}^{r}_{x}.$

           It follows from these relations that the representation of
           $\bar{H}_{y}$ on $\bar{H}_{x}$ can be described as the structure,
           \begin{equation}\label{HcxHx}\begin{array}{l}\bar{H}^{r}_{x}
           =\{H_{x},\pm^{r}_{x},\cdot^{r}_{x},\langle-,-\rangle^{r}_{x},
           \psi^{r}_{x}\}\\\hspace{1cm}=\{H_{x},\pm_{x},\frac{\textstyle
           \cdot_{x}}{\textstyle r},\frac{\textstyle\langle -, -\rangle_{x}}
           {\textstyle r},rV_{y,x}\psi_{x}\}.\end{array}\end{equation}
           The first line in the equation is a repetition of Eq. \ref{Hcyxx}.
           The second  line in Eq. \ref{HcxHx} gives a representation
           of the elements of the structure $\bar{H}^{r}_{x}$ in terms
           of elements of $\bar{H}_{x}.$  It is also part of a local
           representation of $\bar{H}_{y},\bar{C}_{y}$ on
           $\bar{H}_{x},\bar{C}_{x}.$ Thus $rV_{y,x}\psi_{x},
           \langle-,-\rangle_{x}/r, \cdot_{x}/r$ are representations
           of $\psi^{r}_{x},\langle -,-\rangle^{r}_{x},$
           $\cdot^{r}_{x}$ in $\bar{H}_{x}$ and local representations
           of $\psi_{y},\langle-,-\rangle_{y},\cdot_{y}$ in $\bar{H}_{x}.$

           The blanks in $\langle-,-\rangle^{r}_{x}$ denote
           vectors $\psi^{r}_{x},\phi^{r}_{x}$, and the blanks in
           $\langle -, -\rangle_{x}$ denote vectors
           $rV_{y,x}\psi_{x},rV_{y,x}\phi_{x}.$ Eq. \ref{xrxx} was
           used to obtain the relation between the scalar products
           in the two representations of $\bar{H}^{r}_{x}.$

           \section{Gauge Theories}\label{GT}

           The material presented so far forms a base for further
           explorations into the possible effect of the gauge
           field, $\vec{A}(x),$ in physics and mathematics. One
           direction to explore is the description of gauge theories,
           which have been so important in physics \cite{Yang,Novaes,Cheng}.
           Here the discussion is limited to very elementary aspects of
           these theories with their associated Lagrangians. Because of
           the presence of different scalar structures at each space
           time point, the discussion is a bit more detailed than would
           otherwise be needed.

           Let $\psi$ be a field  such that, for each point $x,$ $\psi(x)$
           is a vector in an $n$ dimensional Hilbert space,
           $\bar{H}_{x}.$\footnote{The description given here applies to
           other vector spaces. The choice of Hilbert spaces as examples
           of vector spaces is simply to be able to work with a specific
           and well known example.}  Relative to basis
           choices in each $\bar{H}_{x},$ $\psi$ is an $n$ component complex scalar
           field. Since  $n$ dimensional Hilbert spaces can be represented
           by  $n$ tuples, $\bar{C}^{n},$ of complex number fields \cite{Kadison},
           $\psi(x)$ can be thought of as an element of  $\bar{C}^{n}_{x}.$

           The dynamics of the fields are described by  Lagrangians that include
           terms containing space and time derivatives of the fields.
           Examples include the Klein Gordon and Dirac Lagrangians as
           \begin{equation}\label{KleinG}\mathcal{L}(x)=\psi^{\dag}(x)
           \partial^{\mu}_{x}\partial_{\mu,x}\psi -m^{2}\bar{\psi}(x)\psi(x)
           \end{equation} and\begin{equation}\label{Dirac}\mathcal{L}(x)=
           \bar{\psi}(x)i\gamma^{\mu}\partial_{\mu,x}\psi-m\bar{\psi}(x)\psi(x).
           \end{equation} In these expressions the derivative $\partial_{\mu,x}$
           is given by Eq. \ref{dmux}, which is repeated here,
           \begin{equation}\label{Deltamux}\partial_{\mu,x}\psi=
           \frac{\psi(x+dx^{\mu})-\psi(x)}{\partial x^{\mu}}.
           \end{equation}

           As noted for Eq. \ref{dmux},  the derivative in Eq. \ref{Deltamux}
           is not defined. The reasons are that subtraction is not defined
           between elements of $\bar{H}_{x}$ and $\bar{H}_{x+dx^{\mu}}$ and the
           "no information at a distance principle" prevents an
           observer at $x$ from direct access to vectors at $x+dx^{\mu}.$

           This problem is solved  by replacing $\psi(x+dx^{\mu})$
           with the vector in $\bar{H}_{x}$ that corresponds to the
           vector $\psi(x+dx^{\mu})^{r}_{x}$ in $\bar{H}^{r}_{x}.$
           Recall that $\bar{H}^{r}_{x}$ is the local representation
           of $\bar{H}_{x+dx^{\mu}}$ on $\bar{H}_{x}.$

           The replacement vector is given by\begin{equation}
           \label{mfrkpsi}\begin{array}{l}(\mathfrak{V}^{
           \mu}_{r})^{-1}\psi(x+dx^{\mu})=\mathfrak{V}^{r}_{x}
           V_{\mu,x}\psi(x+dx^{\mu})_{x}\\\\\hspace{1cm}=e^{A_{\mu}(x)
           dx^{\mu}}V_{\mu,x}\psi(x+dx^{\mu})_{x}.
           \end{array}\end{equation} Here $\psi(x+dx^{\mu})_{x}=
           (U_{\mu,x})^{-1}\psi(x+dx^{\mu})$ where the
           parallel transform operator, $U_{\mu,x},$ from $\bar{H}_{x}$ to
           $\bar{H}_{x+dx^{\mu}}$ is given by Eqs. \ref{HyMfrkVHx} and
           \ref{3stepcyx}. To save on notation, $x+dx^{\mu}$ is often
           replaced by $\mu$ as in $\mathfrak{V}^{\mu}_{x}.$ Also
           $r=r_{\mu,x}$ is the $\mu$ component of $r_{y,x}.$

           The replacement vector can also be
           described as the vector value, in $\bar{H}_{x},$ of the
           element of the base set, $H_{x},$ that has the value
           $\psi(x+dx^{\mu})^{r}_{x}$ in $\bar{H}^{r}_{x}.$ This is
           to be distinguished from another base set element that
           has the value $\psi(x+dx^{\mu})_{x}$ in $\bar{H}_{x}.$
           This is the same vector value in $\bar{H}_{x}$ as
           $\psi(x+dx^{\mu})^{r}_{x}$ is in $\bar{H}^{r}_{x},$ and as
           $\psi(x+dx^{\mu})$ is in $\bar{H}_{x+dx^{\mu}}.$

           Eq. \ref{mfrkpsi} is used to express the covariant
           derivative, $D_{\mu,x},$ as
           \begin{equation}\label{DelFmux}D_{\mu,x}\psi=
           \frac{\mathfrak{V}^{r}_{x}V_{\mu,x}\psi(x+dx^{\mu})_{x}
           -\psi(x)}{\partial x^{\mu}}\end{equation} as a
           replacement for $\partial_{\mu,x}$ in the Lagrangians.
           Here $\mathfrak{V}^{r}_{x}$ and $V_{\mu,x}$ account,
           respectively, for the freedom of choice of complex
           number structures and of bases in the Hilbert spaces.

           In the following, the consequences of the replacement
           of ordinary derivatives by $D_{\mu,x}$
           in Lagrangians will be described for Abelian and
           nonabelian gauge theory \cite{Montvay,Cheng,Peskin}.
           The gauge group  for Abelian theories is $GL(1,r)\times U(1),$
           and $GL(1,R)\times U(n)$ with $n>1$ for nonabelian theories.
           For Abelian theories, group elements $\mathfrak{V}^{r}_{x}V_{y,x}$
           have Lie algebra representations as\begin{equation}\label{VcVeV}
           \mathfrak{V}^{r_{y,x}}_{x}V_{y,x}=e^{\vec{A}(x)\cdot\hat{\nu}dx}
           e^{i\vec{\Gamma}(x)\cdot\hat{\nu}dx}.\end{equation} For nonabelian
           theories there is an additional factor for the Lie algebra
           representations of elements of $SU(n).$

           \subsection{Abelian Gauge Theory}\label{ABGT}
           For Abelian theories the covariant derivative can be
           expressed by\begin{equation}\label{Delcyx}D_{\mu,x}
           \psi=\frac{e^{A_{\mu}(x)dx^{\mu}}e^{i\Gamma_{\mu}(x)
           dx^{\mu}}\psi(x+dx^{\mu})_{x}-\psi(x)}{\partial x^{\mu}}.
           \end{equation}

           Expansion of the exponential in $D_{\mu,x}$ to first order in
           small quantities gives the relation between
           $\partial^{\prime}_{\mu,x},$ defined in Eq.
           \ref{parprim}, and $D_{\mu,x}.$ This is
           \begin{equation}\label{Codd}D_{\mu,x}\psi
           =(\partial^{\prime}_{\mu,x}+g_{R}A_{\mu}(x)
           +ig_{I}\Gamma_{\mu}(x))\psi(x).\end{equation}Coupling
           constants for $\vec{A}(x)$ and $\vec{\Gamma}(x)$ have
           been included.

           A well known requirement on  a Lagrangian is that each
           term must be invariant under both global and local gauge
           transformations. Global transformations have the form $\Lambda_{x}=
           e^{i\phi_{x}}$ where $\phi_{x}$ is the same number in
           $\bar{C}_{x}$ as $\phi_{y}$ is in $\bar{C}_{y}.$

           Local gauge transformations, $\Lambda(x)=e^{i\phi(x)},$
           satisfy\begin{equation}\label{Fyxloc} F_{y,x}\Lambda(x)
           =(\Lambda(x))_{y}=(e^{i\phi(x)})_{y}=e^{i\phi(x)_{y}}
           \neq\Lambda(y)=e^{i\phi(y)}.\end{equation}  Here
           $\Lambda(x)$ is different for different $x$ as
           $\phi(x)_{y}\neq \phi(y).$

           Terms in the Lagrangian containing the covariant
           derivative are invariant under global gauge transformations
           if $D_{\mu,x}\Lambda\psi=\Lambda_{x}D_{\mu,x}\psi.$ This
           follows from the observation that $\partial^{\prime}_{\mu,x}
           \Lambda_{x}=0.$ Terms are invariant under local transformations
           if \begin{equation}\label{loclambpsi}D^{\prime}_{\mu,x}
           \Lambda\psi=\Lambda(x)D_{\mu,x}\psi.\end{equation}

           Here  $D^{\prime}_{\mu,x}\Lambda\psi$ is obtained from Eq. \ref{Codd}
           as\begin{equation}\label{MFRKDVG}D^{\prime}_{\mu,x}\Lambda
           \psi=\partial^{\prime}_{\mu,x}\Lambda\psi+(g_{R}A^{\prime}_{\mu}(x)
           +ig_{I}\Gamma^{\prime}_{\mu}(x)) \Lambda(x)\psi(x).\end{equation}
           This equation and Eq. \ref{loclambpsi} are used in the standard
            procedure for Abelian gauge theories \cite{Cheng} to give
            \begin{equation}\label{ApBpAB}\begin{array}{l} (g_{R}A'_{\mu}(x)+
            ig_{I}\Gamma^{\prime}_{\mu}(x))\Lambda(x)\psi(x)\\\\
            \hspace{1cm}=\Lambda(x)(g_{R}A_{\mu}(x)+ig_{I}\Gamma_{\mu}(x))\psi(x)
            -\partial^{\prime}_{\mu,x}(\Lambda)\psi(x).\end{array}
            \end{equation}Use of $\partial^{\prime}_{\mu,x}
            (\Lambda)=i\partial^{\prime}_{\mu,x}
            (\phi(x))\Lambda(x)$ and separation of Eq. \ref{ApBpAB}
            into two separate equations for the real and imaginary
            parts gives the result that \begin{equation}\label{AR}
            A^{\prime}_{\mu}(x)=A_{\mu}(x)\end{equation} and
            \begin{equation}\label{BI}\Gamma^{\prime}_{\mu}(x)=\Gamma_{\mu}(x)+
            \frac{i\Lambda^{-1}(x)\partial^{\prime}_{\mu,x}\Lambda}{g_{I}}=
            \Gamma_{\mu}(x)-\frac{1}{g_{I}}\partial^{\prime}_{\mu,x}\phi(x).
            \end{equation}

          This result shows that the effect of local gauge transformations
          is limited to the gauge field $\Gamma_{\mu}(x)$ as $A_{\mu}(x)$  is
          unaffected.  That is, $A_{\mu}(x)$ is gauge invariant.
          It follows that  $\vec{A}(x)$ and $\vec{\Gamma}(x),$ correspond
          respectively to two gauge bosons, one for which having mass is
          possible, and the other which must be massless.

          The dynamics of the massless boson can be added to the Lagrangian in
          the standard way by addition of a gauge invariant Yang Mills term,
           \begin{equation}\label{Gmnnm}-\frac{1}{4}G_{I,\mu,\nu}G^{\mu,\nu}_{I}
          \end{equation} for $\Gamma_{\mu}(x).$ Here
          \begin{equation}\label{Gmunu}G_{I,\mu,\nu} =\partial^{\prime}_{
          \mu,x}\Gamma_{\nu}(x)- \partial^{\prime}_{\nu,x}\Gamma_{\mu}(x).
          \end{equation}

          Addition of the term of Eq. \ref{Gmnnm} and a mass term for
          the field, $A_{R,\mu}(x),$ in the Dirac Lagrangian gives
          \begin{equation}\label{Diracexp}\begin{array}{l}\mathcal{L}(x)=
          \bar{\psi}i\gamma^{\mu}(\partial^{\prime}_{\mu,x}+g_{R}A_{\mu}(x)
          +ig_{I}\Gamma_{\mu}(x))\psi-m\bar{\psi}\psi\\\\\hspace{1cm}-
          \frac{1}{2}\lambda^{2}A^{\mu}(x)A_{\mu}(x)-\frac{1}{4}
          G_{I,\mu,\nu}G^{\mu,\nu}_{I}.\end{array}\end{equation}
          Except for the terms involving $A_{\mu}(x),$  this has
          the same form as the Lagrangian for QED. This shows that,
          for this setup, the QED Lagrangian, is obtained by setting
          $\vec{A}(x)=0$ for  all $x.$

          \subsection{Nonabelian Gauge Theory}\label{NAGT}
          Here the simplest case for a nonabelian gauge theory is
          considered. Let $\psi$ be a two dimensional field  where
          for each $x,$ $\psi(x)$ is a vector in a two dimensional
          Hilbert space $\bar{H}_{x}.$ Relative to bases in the spaces
          $\bar{H}_{x},$ $\psi$ is a two dimensional complex scalar
          field.

          The Dirac and Klein Gordon Lagrangians have the  form
          as shown in Eqs. \ref{KleinG} and \ref{Dirac} with
          $D_{\mu,x}$ replacing $\partial_{\mu,x}.$
          For each $x$ the scalar product\begin{equation}
          \label{scprod}\bar{\psi}(x)\cdot\psi(x)=
          \bar{\psi}(x)^{1}\psi(x)^{1}+\bar{\psi}(x)^{2}\psi(x)^{2}
          \end{equation} is a number in $\bar{C}_{x}.$ As was the
          case for Abelian gauge theory $D_{\mu,x}$ is given by Eq.
          \ref{DelFmux}. However, $V_{\mu,x},$ as an element of $U(2),$ is given
          by\begin{equation}\label{VmuSU2}V_{\mu,x}=e^{i\Gamma_{\mu}
          (x)dx^{\mu}}e^{-i\vec{\Omega}_{\mu}(x)\cdot\frac{\tilde{\tau}}{2}
          dx^{\mu}}.\end{equation}

          Here\begin{equation}\label{Omtaujj}
          \vec{\Omega}_{\mu}(x)\cdot\tilde{\tau}=\Omega_{\mu}^{j}(x)\tau_{j}
          \end{equation} where the $j$ indices are summed over.
          $\vec{\Omega}_{\mu}$ is a three component vector gauge
          field whose components, $\Omega_{\mu}^{j},$ represent the
          three vector gauge bosons, and the $\tau_{j}$ are the
          generators of the Lie algebra $su(2).$ As  Pauli spin
          operators, the $\tau_{j}$ satisfy the commutation rule,
          \begin{equation}\label{commpauli}[\frac{\tau_{j}}{2},
          \frac{\tau_{k}}{2}]=i\xi_{jkl}\frac{\tau_{l}}{2}
          \hspace{2cm}j,k,l=1,2,3.\end{equation} The structure constant,
          $\xi_{jkl},$ is antisymmetric under exchange of indices.
          At each point $x$ the  vector components,
          $\Omega_{\mu}^{j}(x),$  are real numbers in $\bar{C}_{x}$ and
          the elements of the Pauli matrices are real or imaginary
          numbers in $\bar{C}_{x}.$

          The gauge field representation of the product
          $\mathfrak{V}^{\mu}_{r}V_{\mu,x}$ is given by\begin{equation}
          \label{cmuxVmux}\mathfrak{V}^{\mu}_{r}V_{\mu,x}=e^{A_{\mu}(x)
          dx^{\mu}}e^{i\Gamma_{\mu}(x)
          dx^{\mu}}e^{-i\vec{\Omega}_{\mu}(x)\cdot\frac{\tilde
          {\tau}}{2}dx^{\mu}}.\end{equation} Here the $A_{\mu}(x)$
          are the components of the  gauge field defined by Eq.
          \ref{psicxMFK}. Expansion of the exponentials and retention of
          terms to first order in small quantities gives
          a generalization of Eq. \ref{Codd}:\begin{equation}
           \label{DDAAOm}D_{\mu,x}\psi=(\partial^{\prime}_{\mu,x}+g_{R}
           A_{\mu}(x)+ig_{I}\Gamma_{\mu}(x)-ig\vec{\Omega}_{\mu}(x)
           \cdot\tilde{\tau})\psi(x).\end{equation} Here $g$ is
           the coupling constant for $\vec{\Omega}.$

           The requirement that the Lagrangians be invariant under
           local $U(2)$ gauge transformations is expressed by
           \begin{equation}\label{LambdTh}\Lambda(x)=e^{i\phi(x)}
           e^{-i\vec{\Theta}(x)\cdot\tilde{\tau}/2}=\Lambda_{1}(x)
           \Lambda_{2}(x),\end{equation} Use of Eqs. \ref{loclambpsi}
           and \ref{DDAAOm}, and the commutativity of $\vec{A}$ and
           $\vec{\Gamma}$ with $\vec{\Theta}\cdot\tau$
           gives \cite{Cheng} the result that\begin{equation}\label{ApApOmp}
           \begin{array}{l}(g_{R}A'_{\mu}(x)+ig_{I}\Gamma'_{\mu}(x)-ig\vec{
           \Omega}'_{\mu}(x)\cdot\tilde{\tau})\Lambda_{1}(x)\Lambda_{2}(x)
           \\\\\hspace{1cm}=\Lambda_{1}(x)\Lambda_{2}(x)(g_{R}A_{\mu}(x)
           +ig_{I}\Gamma_{\mu}(x)-ig\Lambda(x)\vec{\Omega}_{\mu}(x)\cdot
           \tilde{\tau})\\\\\hspace{2cm}-\partial^{\prime}_{\mu,x}
           (\Lambda_{1}(x))\Lambda_{2}(x)-\partial^{\prime}_{\mu,x}
           (\Lambda_{2}(x))\Lambda_{1}(x).\end{array}\end{equation}

           This equation has three type of terms, real scalars,
           imaginary scalars, and terms involving the Pauli
           operators. As these are different mathematical types they
           can separately be set equal to $0.$ Since $[\Lambda_{1}(x),
           \Lambda_{2}(x)]=0,$ one
           obtains,\begin{equation}\begin{array}{l}
           \label{ApApRI}A'_{\mu}(x)=A_{\mu}(x),\\\\
          \Gamma'_{\mu}(x)=\Gamma_{\mu}(x)+\frac{i}{g_{I}}
           \partial^{\prime}_{\mu,x}(\Lambda_{1})
           \Lambda^{-1}_{1}(x),\end{array}\end{equation}
           and \cite{Cheng}\begin{equation}\label{OmpOm}
           \vec{\Omega}'_{\mu}(x)\cdot\tilde{\tau}=\Lambda_{2}(x)
           (\vec{\Omega}_{\mu}(x)\cdot\tilde{\tau})\Lambda^{-1}_{2}(x)
           -\frac{i}{g}\partial^{\prime}_{\mu,x}(\Lambda_{2})
           \Lambda^{-1}_{2}(x).\end{equation}

           The Lagrangians are constructed using Eq. \ref{DDAAOm}
           to replace $\partial_{\mu,x}.$ They differ from the
           usual  nonabelian  gauge theory by the presence of
           $A_{\mu}.$ Eq. \ref{ApApRI} shows that, as
           was the case for Abelian gauge theory,  $\vec{A}(x)$
           and $\vec{\Gamma}(x)$ correspond respectively to gauge
           bosons, one for which mass is possible, and the other
           without mass.

           The definition of $D_{\mu,x}$  and its
           use to replace $\partial_{\mu,x}$ in Lagrangians
           should be valid for other gauge groups \cite{Utiyama}.
           For groups such as $GL(1,R)\times U(n)$,
           invariance under local $U(n)$ gauge transformations
           gives the same  results for $\vec{A}(x)$ and
           $\vec{\Gamma}(x)$ as are obtained for the
           nonabelian example described above. However the results
           for $SU(2)$ are replaced by those for $SU(n).$

           \section{Discussion}\label{D}
           There are several open questions associated with the results
           obtained in this paper. Perhaps the most important one
           is concerned with what physical entity is  represented by $\vec{A}(x).$

           This problem does not exist for the gauge field $\vec{\Gamma}(x)$
           in that it must be the photon field. This follows from the observation
           that, if one sets $\vec{A}_{R}=0,$ then the Dirac Lagrangian
           plus the Yang Mills term becomes the usual QED Lagrangian.

           One property that may help to determine the physical nature,
           if any, of $\vec{A}(x)$  is that the ratio of the  $\vec{A}(x)$
           matter field coupling constant, $g_{R},$ to the fine structure
           constant must be very small. This is based on the great accuracy
           of the QED Lagrangian where $\vec{A}(x)$ is absent.

           One may hope that this property can help to determine what
           physical field is represented by $\vec{A}(x).$
           Candidates include the Higgs boson, dark matter,
           dark energy, gravity, and the inflaton \cite{Linde,Albrecht}.
           The small coupling constant requirement suggests that
           $\vec{A}(x)$ may be associated with gravity.
           Dark matter and dark energy cannot be ruled out.
           More work is clearly needed here.

           In 1918 Weyl \cite{Weyl1}, in an attempt to unify
           electromagnetism and gravity, introduced  the condition that the scalar
           product of two vectors at a point  $P$ in Reimannian
           geometry,  is related to the scalar product of these two vectors parallel
           transformed to a neighbor point $P'$ by a scale change factor multiplying the
           metric tensor $g_{i,j}\rightarrow \gamma g_{i,j}.$ If $P$ is at $x$ and $P'$
           is at $x+dx^{\mu}$ and $h$ is any function of space time, then the change
           in $h$ in going from $x$ to $x+dx^{\mu}$ is given  by \cite{Yang1}
           \begin{equation}\label{scch}h\rightarrow h'=h(1+\phi_{\mu}dx^{\mu})
           +(\partial h/\partial x^{\mu})dx^{\mu}.\end{equation} The scale
           change factor is $1+\phi_{\mu}dx^{\mu}.$

           A speculative possibility is that $A_{\mu}(x)=\phi_{\mu}(x).$ Here, unlike
           the case with Weyl's attempt, \cite{Yang1,OR}, there is no problem with
           electromagnetism in that $\vec{A}(x)$ is not related to the electromagnetic
           field. If $\vec{A}(x)$ is a scale change factor for the metric tensor it
           would imply a deep connection between general relativity and mathematics
           in that $\vec{A}(x)$ is also a space time dependent scale change factor
           for complex number structures. Whether there is any merit in these
           speculations or not will have to await further work.

           Another open problem concerns the integrability of
           $\vec{A}(x).$ It is not known if $\vec{A}(x)$ is integrable
           or not. Nonintegrability of $\vec{A}(x)$ would cause
           problems for integrals of complex valued functions over
           space time in that a path dependence would have to be included.
           (See subsection \ref{STI}.)

           A well known example in physics that would have this
           integrability problem is the action, which is a space
           time integral of the Lagrangian density. If $\vec{A}$
           were nonintegrable, the action would have the form of Eq. \ref{intx}
           with  $\Phi$ replaced by the Lagrangian density. In this case
           the integral would depend on the path $P$ from $x$ to $y.$

           It is fortunate that the integrability of $\vec{A}$ is independent
           of that for $\vec{\Gamma},$ which represents the photon field. As
           is well known from the Aharonov-Bohm effect \cite{Bohm},
           the photon gauge field is nonintegrable.

           In this paper, the treatment of gauge theories where
           separate complex number structures are assigned to each
           space time point has been limited to a real gauge field.
           Here the local representation of $\bar{C}_{y}$ on
           $\bar{C}_{x}$ is described in terms of a real number
           $r_{y,x}$ where\begin{equation}\label{ryxA}
           r_{y,x}=e^{\vec{A}(x)\cdot\hat{\nu}dx}.\end{equation}
           This can be expanded by replacing $r_{y,x}$ by a complex
           number $c_{y,x}$ and letting $\vec{A}(x)$ be a complex
           valued gauge field as in $\vec{A}(x)=\vec{A}_{R}(x)+
           i\vec{A}_{I}(x).$

           This gives a more complex theory, both in terms of the
           relation of the local representation of $\bar{C}_{y}$ on
           $\bar{C}_{x}$ and in terms of the fields entering in the
           covariant derivatives for the Lagrangians.  The various
           complications, which result from the generalization,
           seem manageable. This will be shown
           in a separate paper.

           An interesting future direction of work is to expand the space time
           dependence of complex number structures to include other mathematical
           systems, besides Hilbert spaces, that are based on real or complex
           scalar fields. Presumably this includes much of the mathematics used
           by theoretical physics.

           These extensions can be used as a possible approach to a coherent
           theory of physics and mathematics together
           \cite{Tegmark,BenTCTPM,BenTaCTPM}. In this approach, the
           mathematics available to an observer is available locally at each point of
            a world line which is the observer's path through space time.
            One must show that the resulting space time dependence of scalar
            field based mathematics does not introduce inconsistencies for
            comparisons between theoretical and experimental results at
            different points. These ideas will be developed in future work.

            Whatever one thinks of the work presented here, it is worth
            emphasizing again that this work generalizes the existing
           treatment of gauge theories by introduction of the freedom
           of complex number structure choice, in addition to the
           usual freedom of basis choice in the Hilbert spaces. The
           usual setup, with one complex number field for all space
           time points, is obtained by setting the gauge field
           $\vec{A}(x)=0.$

          \section*{Acknowledgement}
          This work was supported by the U.S. Department of Energy,
          Office of Nuclear Physics, under Contract No.
          DE-AC02-06CH11357.

          \end{document}